\documentclass[12pt] {iopart}

 \expandafter\let\csname equation*\endcsname\relax
  \expandafter\let\csname endequation*\endcsname\relax
  
\usepackage{amsmath}
\usepackage{iopams}  

\begin{document}

\title[A Grasp of Identity]{A Grasp of Identity}

\author{Pere Seglar and Enric P\'{e}rez}

\address{Universitat de Barcelona, Departament de F\'isica Fonamental, c. Mart\'i i Franqu\`es 1, Barcelona 08028}
\ead{enperez@ub.edu}

\begin{abstract}
We revisit the treatment of identical particles in quantum mechanics. Two kinds of solutions of Schr\"{o}dinger equation are found and analyzed. First, the known symmetrized and antisymmetrized eigenfunctions. We examine how the very concept of particle is blurred whithin this approach. Second, we propose another kind of solution with no symmetries that we identify with Maxwell-Boltzmann statistics. In it, particles do preserve their individuality, as they are provided with individual energy and momenta. However, these properties cannot be univocally ascribed; moreover, particles do not possess distinctive positions. Finally, we explore how these results affect the calculation of canonical partition function, and we show that extensivity arises as a consequence of identity.

\end{abstract}

\maketitle

\section{Introduction}

Elementary particles of the same kind are identical: their elementarity entails the absence of intrinsic differences among them. In classical mechanics this does not represent a problem. It does only when we move to statistical mechanics. Whereas in classical mechanics variable $N$ has ordinal and cardinal meaning (it represents the label of the $N$th particle as well as the total sum of particles), in thermodynamics this is not the case: $N$ has only the cardinal meaning.\footnote{In other words, we can only permute particle indexes in mechanics, not in thermodynamics.} That is the reason why in classical statistical mechanics one has to operate carefully with $N$. In fact, how to achieve extensive thermodynamic potentials within the frame of canonical and microcanonical ensembles has been one of the most discussed issues since the birth of statistical mechanics \cite{mon}. In a recent paper, we have shown that even resorting to quantum mechanics and performing the classical limit, a factor $N^N e^{aN}$ ($a$ is an indeterminate constant) must be added to the canonical partition function in order to obtain an extensive free energy \cite{Seglar}. We argued that the introduction of this factor has nothing to do with \textit{indistinguishability}, as it is usually stated, but with \textit{extensivity}. In the present paper we want to show that it is precisely a proper treatment of \textit{identity} (not of \textit{indistinguishability}) what yields \textit{extensivity}. A proper consideration of the identity of particles within the frame of quantum mechanics leads to extensive thermodynamical potentials, and provides a reasonable justification for the addition of the aforementioned $N$-factor in the canonical partition function.

The classical strategy to distinguish identical particles is through position and momenta, that is, through initial conditions. In contrast, according to the uncertainty principle, in quantum mechanics this cannot be the case. And because of that, the problem of identity arises before moving to statistical mechanics. It arises whithin the frame of quantum mechanics. In Dirac's words \cite{Dirac}:

\begin{quote}
If a system in atomic physics contains a number of particles of the same kind, e.g. a number of electrons, the particles are absolutely indistinguishable one from another. No observable change is made when two of them are interchanged. This circumstance gives rise to some curious phenomena in quantum mechanics having no analogue in classical theory, which arise from the fact that in quantum mechanics a transition may occur resulting in merely the interchange of two similar particles, which transition then could not be detected by any observational means. A satisfactory theory ought, of course, to count two observationally indistinguishable states as the same state and to deny that any transition does occur when two similar particles exchange places. 
\end{quote}

\vspace{1ex}
Still, even if particles are `absolutely indistinguishable,' we are forced to label them $(1 \cdots N)$ to perform calculations. And this is what must be done carefully. We will show that there are two strategies which correspond to two different kind of solutions of Schr\"{o}dinger equation for a system of $N$ identical particles. On the one hand we will be forced to abandon --paradoxically-- any vestige of particles. Whithin this approach we label states, not particles. We will call this procedure \textit{Quantum Identity}, because it consists of the construction of the usual symmetrized and antisymmetrized solutions (Section 3). 

	We will also develop another strategy, based on another solution of Schr\"{o}dinger equation, with which particles do preserve some kind of \textit{individuality}: we label particles by assigning them different momenta. However, particles are not localized at all. We will call this aproach \textit{Semiclassical Identity} (Section 4). 

Although our arguments mainly lie in mathematical results, they easily lead to fundamental questions related to the status of atomism according to quantum mechanics. As we will see, the meaning of symmetrization of eigenfunctions is closely related to wave/particle dualism and entanglement. More specifically, we think our reflections can throw new light on recent (and early) debates on the actual status of particles in quantum mechanics \cite{schro, hob, sci}. Hence, our paper provides new insights on an old subject, and we hope it will be interesting for teachers and students of statistical mechanics and quantum mechanics. Any general physicist should be able to read it, as well as any graduate student.

\section{Identical Particles}

\subsection{Classical labels}

As we have said, if two particles are identical, they possess no features which make possible to distinguish them. Only external factors do. The hamiltonian of an $N$-identical free particle system is symmetric, that is, it does not distinguish among particles:

$$H=\sum_{i=1}^{N} \frac{\vec{p}_{i}}{2m}.$$

Position is the factor employed in the frame of classical mechanics to distinguish them, because the dynamics of the system never interchanges particles:

$$\dot{q}_{k}=  \frac{\partial H}{\partial {p_k}}  \qquad \dot{p}_{k}=  - \frac{\partial H}{\partial {q_k}}$$
($q_{k}$ are generalized position coordinates, and $p_{k}$ momenta). Hence, in order to label particles, we must use initial conditions, and pass from the so-called $\Gamma$-space (the system space) to the $\mu$-space (individual space) \cite{ehr}.\footnote{Of course, there is some aribitrariness in assigning indexes to particles. However, different stipulations will yield the same dynamics. Moreover, we perfectly know how to pass from one to another. Recall we are still in the frame of mechanics, not statistical mechanics.} As far as we have $N$ points in the $\mu$-space at the intial instant, identical particles are unequivocally labelled. 

\subsection{Quantum labels}

Within the frame of quantum mechanics we cannot appeal to initial conditions nor to $\mu$-space, as we cannot make use of phase coordinates. What we can surely say is that were particles identical, all observables $\hat{\Theta}$ would be symmetric:

\begin{equation}
\hat{\Theta}(1\cdots i, j \cdots N)=\hat{\Theta}(1\cdots j,i \cdots N).
\end{equation}
In general:

$$[\hat{\Theta} (1 \cdots N), \hat{P}]=0,$$
where with $\hat{P}$ we indicate any of the $N!$ possible permutations of indexes $(1 \cdots N)$. As for the hamiltonian, the condition reads:

\begin{equation} \label{hamisimi}
[\hat{H}(1 \cdots N), \hat{P}]=0.
\end{equation}
The subject of the present paper are the awkward consequences of this condition: the so-called \textit{interchange degeneration}. This interchange refers to the indexes $(1 \cdots N)$: for the same physical system different (and orthogonal) solutions of the same Schr\"{o}dinger equation are possible.

Let us analyze the equation:

\begin{equation}
 \hat{H}\psi_{E_{\alpha}}(1 \cdots N)=E_{\alpha}\psi_{E_{\alpha}}(1 \cdots N).
\end{equation}
In general, this solution has degeneration $N!$ That is, starting from the solution $\psi_{E_{\alpha}}(1 \cdots N)$ we can obtain additional solutions just applying the permutation operator:

\begin{equation}
\hat{P}_{ij} \psi_{E_{\alpha}}(1 \cdots N)
\end{equation} 
($P_{ij}$ indicates permutation between indexes $i$ and $j$). These solutions are different, but refer to the same physical system. What we have to do is supressing this \textit{degeneration due to indexation}. First of all, we must remember that eigenfunctions themselves are not observable. That is, for all observational purposes, $\psi_{E_{\alpha}}(1 \cdots N)$ and $\psi_{E_{\alpha}}(1 \cdots N)e^{i \theta}$ are identical. Only the square of the wave function can be directly related to observations; only in the squared function the identity can be imposed:

\begin{equation} \label{id}
|\psi_{E_{\alpha}}(1 \cdots i,j \cdots N)|^{2}\equiv|\psi_{E_{\alpha}}(1 \cdots j,i \cdots N)|^{2}.
\end{equation}
This condition has three solutions which correspond to three different ways of imposing the elimination of interchange degeneration:

\begin{equation} \label{mingulis}
\begin{array}{l}
\mbox{A} \qquad  \psi_{E_{\alpha}}(1 \cdots i,j \cdots N)=\psi_{E_{\alpha}}(1,2\cdots j,i \cdots N)\\
\mbox{B} \qquad  \psi_{E_{\alpha}}(1 \cdots i, j \cdots N)=-\psi_{E_{\alpha}}(1,2\cdots j,i \cdots N)\\
\mbox{C} \qquad  {|\psi_{E_{\alpha}}(1 \cdots i,j \cdots N)|}^{2}=const. 
\end{array}
\end{equation}
A and B correspond to the known symmetric and antisymmetric solutions (\textit{Quantum Identity}). C, as we will argue, corresponds to the Maxwell-Boltzmann solution within the frame of quantum mechanics (\textit{Semiclassical Identity}). In the following, we will show that these solutions exist, are unique, and have physical meaning.  

\section{Quantum Identity}

Let us begin with the case $N=2$. As it is known \cite{gott}:

\begin{equation} \label{simio}
\psi_{E_{\alpha}}^{S/A}(1,2)\equiv \frac{1}{\sqrt{2}} [\psi_{E_{\alpha}}(1,2)\pm \psi_{E_{\alpha}} (2,1)]
\end{equation}
(the superindex indicates symmetrized/antisymmetrized solutions).\footnote{We distinguish \textit{symmetric} and \textit{symmetrized} eigenfunctions. The previous are originally symmetric, as solutions of Schr\"{o}dinger equation. The latter are constructed through the known rule \eqref{psisa}. See \cite{Seglar}.}  Eigenfunctions \eqref{simio} have no interchange degeneration: their square remains the same when we permute indexes $1$ and $2$. In this case, we can easily invert the relation:

\begin{equation} \label{pemmuta}
\psi_{E_{\alpha}}(1,2)\equiv \frac{1}{\sqrt{2}} [\psi^{S}_{E_{\alpha}}(1,2)+ \psi^{A}_{E_{\alpha}} (2,1)].
\end{equation}
Hence, with symmetrized and antisymmetrized functions  we can construct the whole subspace of solutions with energy $E_{\alpha}$. This is related to the fact that the  two only possible permutations commute. In this sense, $N=2$ is not a good example of the problem we are dealing with. $N=3$ is the first non trivial case. Now:\footnote{When the two pairs of indexes are different permutations do commute.}

\begin{equation}
[\hat{P}_{ij}, \hat{P}_{jk}]\neq 0.
\end{equation}
Nevertheless, there are common eigenfunctions to all possible permutations.\footnote{This is the case because $\hat{P}_{ij}$ is not an observable.} Consider eigenfunction $\psi_{E_{\alpha}}(1,2,3)$. The possible permutations are $6=3!$ Now we cannot write any eigenfunction only in terms of symmetrized and antisymmetrized functions. We are forced to add eigenfunctions with other symmetries, which we will call mixed symmetries. These are the corresponding functions for $N=3$ (symmetrized, antisymmetrized, and the rest, with mixed symmetries) \cite{Galindo}: 

\begin{equation} \label{sym}
\begin{array} {c}
\psi_{E_{\alpha}}^{S}(1,2,3)=\frac{1}{\sqrt{6}} \left[\psi_{E_{\alpha}}(1,2,3) +\psi_{E_{\alpha}}(1,3,2)+\psi_{E_{\alpha}}(2,3,1) + \right. \\ \left. +\psi_{E_{\alpha}}(2,1,3)+\psi_{E_{\alpha}}(3,1,2)+\psi_{E_{\alpha}}(3,2,1) \right]
\end{array}
\end {equation}

\begin{equation}  \label{antisym}
\begin{array} {c}
\psi_{E_{\alpha}}^{A}(1,2,3)= \frac{1}{\sqrt{6}} \left[\psi_{E_{\alpha}}(1,3,2) -\psi_{E_{\alpha}}(1,2,3)+\psi_{E_{\alpha}}(2,3,1)- \right. \\ \left. -\psi_{E_{\alpha}}(2,1,3)+\psi_{E_{\alpha}}(3,1,2)-\psi_{E_{\alpha}}(3,2,1) \right]
\end{array}
\end {equation}

\begin{equation} \label{M1}
\begin{array} {c}
\psi_{E_{\alpha}}^{s_1}(1,2,3)= \frac{1}{2\sqrt{3}} \left[2 \psi_{E_{\alpha}}(1,2,3) -\psi_{E_{\alpha}}(1,3,2)+2\psi_{E_{\alpha}}(2,1,3) - \right. \\ \left. - \psi_{E_{\alpha}}(2,3,1) - \psi_{E_{\alpha}}(3,1,2) - \psi_{E_{\alpha}}(3,2,1) \right]
\end{array}
\end {equation}

\begin{equation} \label{M2}
\begin{array} {c}
\psi_{E_{\alpha}}^{s_2}(1,2,3)=\frac{1}{2} \left[\psi_{E_{\alpha}}(1,3,2) -\psi_{E_{\alpha}}(2,3,1)+\psi_{E_{\alpha}}(3,1,2) - \psi_{E_{\alpha}}(3,2,1)\right]
\end{array}
\end {equation}

\begin{equation} \label{M3}
\begin{array} {c}
\psi'^{s_1}_{E_{\alpha}}(1,2,3)= \frac{1}{2\sqrt{3}} \left[2 \psi_{E_{\alpha}}(1,2,3) +\psi_{E_{\alpha}}(1,3,2)-2\psi_{E_{\alpha}}(2,1,3) -\right. \\ \left. - \psi_{E_{\alpha}}(2,3,1) - \psi_{E_{\alpha}}(3,1,2) + \psi_{E_{\alpha}}(3,2,1) \right]
\end{array}
\end {equation}

\begin{equation} \label{M4}
\begin{array} {c}
\psi'^{s_2}_{E_{\alpha}}(1,2,3)= \frac{1}{2} \left[\psi_{E_{\alpha}}(1,3,2) +\psi_{E_{\alpha}}(2,3,1)-\psi_{E_{\alpha}}(3,1,2) - \psi_{E_{\alpha}}(3,2,1)\right].
\end{array}
\end{equation}
The last four eigenfunctions are grouped in pairs, \eqref{M1} and \eqref{M2}, and \eqref{M3} and \eqref{M4}: to generate their own subspace they require two independent functions instead of one. Permutations performed in \eqref{M1} yield a function which is a combination of \eqref{M1} and \eqref{M2}, and the same with \eqref{M2} (\textit{ditto} for \eqref{M3} and \eqref{M4}). But under consecutive permutations the two generated subspaces remind stable. No permutation will transform \eqref{sym} or \eqref{antisym} into any other (that is, symmetrized functions remain symmetric and antisymmetrized functions remain antisymmetric). With a little calculation we can find that we can express $\psi_{E_{\alpha}}$ as follows:

$$\psi_{E_{\alpha}}(1,2,3)=\frac{1}{\sqrt{6}}\left[ \psi^{S}_{E_{\alpha}}(1,2,3) + \psi^{A}_{E_{\alpha}}(1,2,3) \right]+ \frac{2 \sqrt{3}}{6} \left[ \psi^{s_1}_{E_{\alpha}}(1,2,3)+  \psi'^{s_1}_{E_{\alpha}}(1,2,3) \right].$$

Recall our aim is to supress interchange degeneration, as it contradicts the idea of identical particles. If we symmetrize solutions, and  take only eigenfunctions \eqref{sym} or \eqref{antisym}, degeneration disappears. In general:

\begin{equation} \label{psisa}
\psi^{S/A}_{E_{\alpha}}(1, \cdots, N) \equiv \frac{1}{\sqrt{N!}} \sum_{P} (\pm 1)^{P} \psi_{E_{\alpha}} (P[1, \cdots, N]).
\end{equation} 

Note that we use the whole set of $N!$ eigenfunctions to construct the new symmetrized/antisymmetrized eigenfunctions. However, degeneration of an arbitrary solution will not always  have the value $N!$: some of the solutions of the Schr\"{o}dinger equation can be symmetric. But no hamiltonian has all its solutions completely symmetric or antisymmetric, as no set of symmetric or antisymmetric functions only can constitute a complete set (eigenfunctions of the hamiltonian cannot be expressed in terms of symmetric and antisymmetric functions only). Therefore, in the non-symmetric solutions for sure there will be interchange degeneration. In conclusion, we are always forced to symmetrize/antisymmetrize some functions and refuse the original with no symmetries.

In sum, we have showed that solution \eqref{psisa} exists in all cases and it is unique. 

\subsection{Physical Meaning}

Which is its physical meaning? Which is the toll it takes suppressing degeneration? To make the analysis easier, let us consider a system of 3 free particles. The hamiltonian is:
 
 $$\hat{H}=\displaystyle{\sum^{3}_{i=1}} \hat{H}_{i},$$
with

$$\hat{H}_{i}=\frac{\hat{p}^{2}_{i}}{2m}.$$
Solution $\psi_{E_{\alpha}}(1,2,3)$ satisfy:

$$\hat{H} \psi_{E_{\alpha}}(1,2,3) =E_{\alpha} \psi_{E_{\alpha}}(1,2,3).$$
We can construct the global solutions starting from monoparticular solutions, as it is usually done:

\begin{equation} \label{prodkt}
\psi(1,2,3)=\varphi_{\epsilon_1}(1)\varphi_{\epsilon_2}(2)\varphi_{\epsilon_3}(3),
\end{equation}
with
$$\hat{H}_{i} \varphi_{\epsilon_{i}}(i)=\epsilon_{i}\varphi_{\epsilon_{i}}(i).$$
The energy can be written as:

$$E_{\alpha}=\epsilon_{1}+\epsilon_{2}+\epsilon_{3}.$$

In general, solution \eqref{prodkt} is not symmetric nor antisymmetric. If the state $\epsilon_{1}=a$ appears 3 times:

\begin{equation} \label{iki}
\psi_{E_{\alpha}}(1,2,3)=\varphi_{a}(1)\varphi_{a}(2)\varphi_{a}(3),
\end{equation}
the function is trivially symmetric: it will be equivalent to \eqref{sym}. If two levels $\epsilon_{1}=a$ are repeated:

\begin{equation} \label{casiki}
\psi_{E_{\alpha}}(1,2,3)=\varphi_{a}(1)\varphi_{a}(2)\varphi_{b}(3),
\end{equation}
we can obtain 3 different solutions through permutations. It is a lineal combination of one pair of mixed symmetries, \eqref{M1} and \eqref{M2}, or \eqref{M3} and \eqref{M4} (in both cases some permutation let the function unaltered and some not), and the symmetrized eigenfunction \eqref{sym}. We have a subspace of dimension $3$. 

With 3 different states degeneration is 6:

$$\varphi_{a}(1)\varphi_{b}(2)\varphi_{c}(3).$$ 
Now the $6$ eigenfunctions \eqref{sym}, \eqref{antisym}, \eqref{M1}, \eqref{M2}, \eqref{M3} and \eqref{M4} are needed because there is no symmetries.

But only \eqref{sym} and \eqref{antisym} are acceptable solutions for identical particles. If we calculate the energy of particle $i$ according to them we obtain:

\begin{equation} \label{mean}
<\psi^{S/A}_{E_{\alpha}}(1,2,3)|\hat{H}_{i}|\psi^{S/A}_{E_{\alpha}}(1,2,3)>=\frac{2}{6} (\epsilon_{1}+\epsilon_{2}+\epsilon_{3})=\frac{E_{\alpha}}{3}.
\end{equation}
Recall in Fermi case we always have different monoparticular states, and sometimes:

$$\epsilon_{1}\neq\epsilon_{2}\neq\epsilon_{3},$$
and for bosons this can also be the case, although not necessarily. However, in both cases, we are forced to say that solutions \eqref{sym} and \eqref{antisym} prevents us to assign particular energies: at the most, according to \eqref{mean}, we can say that each particle has the mean energy. Individuality, as far as energy concerns, has disappeared. 

It is illustrative to see what would have happened had we taken mixed symmetries \eqref{M1}, \eqref{M2}, \eqref{M3} and \eqref{M4}. Let us calculate:\footnote{Note that now we cannot calculate $<H_{i}>$, but only a definite $i=1,2,$ or $3$. This is a consequence of the non-symmetric character of eigenfunctions \eqref{M1}, \eqref{M2}, \eqref{M3} and \eqref{M4}.}

$$<\psi^{s_{1}}_{E_{\alpha}}(1,2,3) |\hat{H}_{1}| \psi^{s_{1}}_{E_{\alpha}}(1,2,3)>=\frac{1}{12} (5\epsilon_{1} + 5 \epsilon_{2} + 2\epsilon_{3})=<\psi^{M_{1}}_{E_{\alpha}}(1,2,3) |\hat{H}_{2}| \psi^{M_{1}}_{E_{\alpha}}(1,2,3)>.$$
And:

$$<\psi^{s_{1}}_{E_{\alpha}}(1,2,3) |\hat{H}_{3}| \psi^{s_{1}}_{E_{\alpha}}(1,2,3)>=\frac{1}{12} (2\epsilon_{1} + 2\epsilon_{2} + 8\epsilon_{3}).$$
That is:

$$<\hat{H}_{1}>_{s_{1}}=<\hat{H}_{2}>_{s_{1}}\neq <\hat{H}_{3}>_{s_{1}},$$
but:

$$<\hat{H}_{1}>_{s_{1}}+<\hat{H}_{2}>_{s_{1}}+ <\hat{H}_{3}>_{s_{1}}=E_{\alpha}.$$
Also:

$$<\psi^{s_{2}}_{E_{\alpha}}(1,2,3) |\hat{H}_{1}| \psi^{s_{2}}_{E_{\alpha}}(1,2,3)>=\frac{1}{4} (\epsilon_{1} + \epsilon_{2} + 2\epsilon_{3})=<\psi^{s_{2}}_{E_{\alpha}}(1,2,3) |\hat{H}_{2}| \psi^{s_{2}}_{E_{\alpha}}(1,2,3)>,$$
and:

$$<\psi^{s_{2}} |\hat{H}_{3}| \psi^{s_{2}}>=\frac{1}{2} (\epsilon_{1} + \epsilon_{2}).$$
That is:

$$<\hat{H}_{1}>_{s_{2}}=<\hat{H}_{2}>_{s_{2}}\neq <\hat{H}_{3}>_{s_{2}},$$
but:

$$<\hat{H}_{1}>_{s_{2}}+<\hat{H}_{2}>_{s_{2}}+ <\hat{H}_{3}>_{s_{2}}=E_{\alpha}.$$
And so on. Then, since particles have different expected values for the energy, mixed symmetries are not acceptable solutions.

Going back to symmetrized/antisymmetrized functions, the same reasoning can be applied in exactly the same way to the expected value of momenta:

\begin{equation} \label{meanp}
<\psi^{S/A}_{E_{\alpha}}(1,2,3)|\hat{\vec{p}}_{i}|\psi^{S/A}_{E_{\alpha}}(1,2,3)>=\frac{<\vec{p}_{1}>+<\vec{p}_{2}>+<\vec{p}_{3}>}{3}.
\end{equation}
Therefore, when using solutions \eqref{sym} and \eqref{antisym} neither energy nor momenta can be individually ascribed. As for position:

\begin{equation} \label{meanr}
<\psi^{S/A}_{E_{\alpha}}(1,2,3)|\vec{q}_{i}|\psi^{S/A}_{E_{\alpha}}(1,2,3)>
\end{equation}
is the geometric center of the container. But, again, for \textit{every particle}, as all of them have the same available volume. Hence, when we calculate the mean position of particles through symmetrized eigenfunctions we find that they are not localized at all. We can imagine an individual particle with a particular energy and momentum, but only outside the container. In the very moment of introducing it into the volume along with a group of $N$ identical particles, it loses its particularities, its individuality. 

Let us comment now on the consequences of quantum symmetry which affect statistical mechanics.

\subsection{The Fock Space}

To calculate the canonical partition function, we have to perform the sum:

\begin{equation} \label{grk}
Z^{S/A}(T,V,N)=\sum_{\underset{\mbox{\scriptsize  \textit{states only}}}{\mbox{\scriptsize \textit{sym./antisym.}}}} e^{-\beta E_{\alpha}}.
\end{equation}
($\beta$ is $1/kT$, and $k$ Boltzmann's constant). There is no degeneration, as we assume to have symmetrized/antisymmetrized eigenfunctions for every energy $E_{\alpha}$. For a system of free particles we could write:

$$Z^{S/A}(T,V,N)=\sum_{\underset{\mbox{\scriptsize\textit{states}}}{\mbox{\scriptsize \textit{monoparticular}}}} e^{-\beta \epsilon_{1}}e^{-\beta \epsilon_2} \cdots$$
But we cannot perform this sum, as particular energies in the exponentials are not independent of each other. We must resort to Fock space \cite{gott}. 

For our purposes, Fock space is a mathematical device where the variables, the so-called \textit{occupation numbers} of monoparticular states $k=a,b,c,...$

\begin{equation} \label{ocupn}
|n_{a}, n_{b}, n_{c}\cdots >
\end{equation}
must satisfy

\begin{equation} \label{seconc}
E=\sum_{k} n_{k} \epsilon_{k}
\end{equation}
and
\begin{equation} \label{ene}
N= \sum_{k} n_{k}.
\end{equation}

We can write $Z^{S/A}(T,V,N)$ as:

\begin{equation} \label{sumac}
\sum_{n_{a}} e^{-\beta n_{a} \epsilon_{a}} \sum_{n_{b}} e^{-\beta n_{b} \epsilon_{b}} \sum_{n_{c}} e^{-\beta n_{c} \epsilon_{c}} \cdots
\end{equation}
However, this sum can neither be performed, because of condition \eqref{ene}. Then, we arrive at a widely known consequence: to calculate the partition function we have to suppress restriction  \eqref{ene}. Only then, when $N$ is not limited, sum \eqref{sumac} can be done, because then $n_{k}$'s are not limited any more. Nevertheless, the sum has lost its physical meaning. To provide it we have to add the chemical potential $\mu$ and move to grand canonical ensemble:

\begin{equation} \label{sumacii}
\Xi (T,V, \mu)=\sum_{n_{a}} e^{-\beta n_{a} (\epsilon_{a}-\mu)} \sum_{n_{b}} e^{-\beta n_{b} (\epsilon_{b}-\mu)} \sum_{n_{c}} e^{-\beta n_{c} (\epsilon_{c} - \mu)} \cdots
\end{equation}

This is the grand partition function. There is an univocal relation between occupation numbers \eqref{ocupn} and symmetrized/antisymmetrized eigenfunctions \eqref{psisa}. According to the usual interpretation, \eqref{ocupn} indicates the number of particles in the states $a$, $b$, etc. On the contrary, strictly speaking, \eqref{ocupn} indicates the number of monoparticular eigenfunctions $\varphi_{a}$, $\varphi_{b}$, etc. which appear in the global eigenfunction. We insist on that no particular state can be ascribed to a definite particle. 

This impossibility of calculating the canonical partition function of bosonic and fermionic gases has a physical origin.	Recall that the thermodynamic potential which we calculate in the canonical ensemble is the Helmholtz free energy, starting from \eqref{grk}:
	
	$$F(T,V,N)=kT \ln Z(T,V,N).$$
Its extensivity allows (and invite) us to write:

\begin{equation} \label{helm}
F(T,V,N)=NF\left(T, \frac{V}{N}, 1 \right).
\end{equation}
This means that the global free energy can be conceived and written as the sum of $N$ individual free energies. This individual free energies can be attributed to $N$ identical components. This interpretation cannot be held when we are dealing with fermionic or bosonic gases. There is nothing like individual particles if we symmetrize the global partition function. This is why canonical (and microcanonical) partition functions of either fermionic and bosonic ideal gases cannot be calculated. 

And what happens in the Fock space? Suppression of condition \eqref{ene} allows us to calculate the sum \eqref{sumac}, but now we have moved to grand canonical ensemble. Our parameter is not $N$ but $\mu$, the chemical potential, and sum \eqref{sumacii} becomes:

\begin{equation} \label{sumacio}
 \prod_{k} \left[ \sum_{n_{k}} \left( e^{-\beta (\epsilon_{k} - \mu)} \right)^{n_{k}} \right].
\end{equation}
The thermodynamic potential does not depend on $N$ any more, but on $\mu$:

$$pV=kT \ln \Xi (T,V, \mu)$$
($p$ is pressure). Now the thermodynamical potential cannot be decomposed into $N$ identical terms and it is trivially extensive: it is always proportional to $V$. However, it is easy to see that no Legendre transformation gives $F$ starting from $\Xi$ with a closed expression (not as an infinite serie). Then, we cannot obtain the (exact) canonical partition function even through the grand canonical.

\section{Semiclassical Identity}

We are going to argue that solution C in \eqref{mingulis} can describe identical particles although it has no symmetry. As we have said, its square is a constant, and in general:

\begin{equation} \label{sotius}
|\psi_{E_{\alpha}}(1 \cdots i,j \cdots N)|=\sqrt{const.} e^{if(1 \cdots i,j \cdots N)}.
\end{equation}

Does it have physical meaning? Consider the hamiltonian of a $N$-particle system:
 
$$H=\sum^{N}_{j=1} - \frac{\hbar^{2}}{2m} i\nabla^{2}_{j}+V(1 \cdots N).$$
In order to be a solution, \eqref{sotius} must fulfil:

\begin{equation}
\hat{H}e^{if(1 \cdots i, j \cdots N)}=E_{\alpha}e^{if(1 \cdots i,j \cdots N)}.
\end{equation}
Now we can say that:
\begin{equation}
\left\{ \left[\sum^{N}_{j=1} - \frac{\hbar^{2}}{2m} i\nabla^{2}_{j} \right] + V(1 \cdots N)  \right\}e^{if(1 \cdots i,j \cdots N)} = E e^{if(1 \cdots i,j \cdots N)}.
\end{equation}
After some calculations:
\begin{equation} \label{casis}
\begin{array} {c}
\displaystyle{\left\{ \sum^{N}_{j=1} -\frac{\hbar^{2}}{2m} \left\{ i\nabla^{2}_{j}f(1\cdots N) -[\nabla_{j} f(1 \cdots N)]^{2} \right\}+ V(1 \cdots N) \right\}} e^{if(1 \cdots i, j \cdots N)} =\\ = E_{\alpha} e^{if(1 \cdots i,j \cdots N)}.
\end{array}
\end{equation}

As in the right side there is no imaginary part and $f$ is real, we can deduce a condition for $f$:

$$\sum_{j} \nabla^{2}_{j} f(1 \cdots N)=0.$$
Although it is the sum what must be null, and not each term, for identical particles differences among indexes must vanish. Hence, in general:

\begin{equation} \label{solitution}
f=\sum^{N}_{j=1} a_{j} \vec{q}_j.
\end{equation}
Therefore, \eqref{casis} becomes:

\begin{equation}
\begin{array} {c}
 \displaystyle{\left\{ \sum^{N}_{j=1} \frac{\hbar^{2}}{2m} \left[\nabla_{j} f(1 \cdots N)\right]^{2} + V(1 \cdots N) \right\}} = E_{\alpha} .
\end{array}
\end{equation}

If now we impose that the left side always should be a number for an arbitrary potential $V(1 \cdots N)$, the potential must be constant. Hence, we can take the value $V(1\cdots N)=0$, which means that we only obtain a solution of this kind for free particles. Then:

\begin{equation}
\begin{array} {c}
 \displaystyle{\sum^{N}_{j=1}} \frac{\hbar^{2}}{2m} [\nabla_{j} f(1 \cdots N)]^{2} = E_{\alpha},
\end{array}
\end{equation}
with:
 
\begin{equation}
\sum^{N}_{j=1} \frac{\hbar^{2}}{2m} a^{2}_{j}=E_{\alpha}.
\end{equation}
 
Finally, if we take $a_{j}=\vec{p}_{j}/h$ we obtain the solution for plane waves as a product of monoparticular eigenfunctions of the form:

\begin{equation} \label{plannn}
\varphi_{j} (\vec{p}_{j}, \vec{q}_{j})=\frac{1}{V^{1/2}} e^{i \frac{\vec{p}_{j} \vec{q}_{j}}{h}}.
\end{equation}
Hence:
\begin{equation} \label{plannes}
\psi_{E_{\alpha}}(1 \cdots N) = \prod_{j} \varphi_{j} (\vec{p}_{j}, \vec{q}_{j})=\frac{1}{V^{N/2}} e^{i \displaystyle{\sum_{j}}\frac{\vec{p}_{j}\vec{q}_{j}}{h}}.
\end{equation}

\subsection{Unicity}

Recall the only thing we have imposed is the identity of particles. In this sense, deduction of solution \eqref{plannes} is spotless. But we can see immediately that there is no univocal correspondence between \eqref{plannes} and the $N$-particle state. There exists a group of different monoparticular products $\vec{p}_{j}\vec{q}_{j}$ compatible with the same physical state $\{\vec{p}_{1}, \vec{p}_{2}, \cdots \}$.\footnote{Note this is not the same problem we mentioned when referring to classical mechanics. There the pairs $\vec{p}_{i}$ and $\vec{q}_{i}$ were linked; now this is not the case.} Again, an interchange degeneration of indexes arises, of value:

\begin{equation}  \label{pitupitu}
\frac{N!}{n_{1}!n_{2}! \cdots n_{j}! \cdots},
\end{equation}
where:
 
$$\sum_{j}n_{j}=N.$$
Here $n_{j}$ means the number of particles with the same momenta $\vec{p}_{j}$. However, to all physical purposes, interchange degeneration of indexes given by \eqref{pitupitu} has no consequences, as their square \eqref{id} is exactly the same. This means that different eigenfunctions yield the same amplitude of probability, as happens with the antisymmetric eigenfunctions, which differ only in a sign. Hence, solution \eqref{plannes} does not deny the existence of particles: there are $N$ (or less) different momenta, with their corresponding monoparticular eigenfunctions, but extended over all the volume. Hence, in contrast to symmetrized and antisymmetrized eigenfunctions (\textit{Quantum Identity}), we now have individual energies and momenta. However, note we cannot ascribe them to certain labeled particles; moreover, particles are not localized. In this sense they are completely \textit{indistinguishable}, even if they possess different momenta.

\subsection{Counting states}

When moving to statistics we have to act carefully. The canonical partition function:

\begin{equation} \label{grk}
Z(T,V,N)=\sum_{\mbox{\scriptsize\textit{states}}} e^{-\beta E_{\alpha}},
\end{equation}
can be calculated through the particular energies:

$$Z(T,V,N)=\sum_{\underset{\mbox{\scriptsize\textit{states}}}{\mbox{\scriptsize \textit{monoparticular}}}} e^{-\beta \epsilon_{1}}e^{-\beta \epsilon_2} \cdots e^{-\beta \epsilon_{N}}.$$
But we have to avoid repetitions in the counting. That is, as far as the $N$ terms are indistinguishable, we cannot extend every variable all over its domain: this would involve an overcounting of states. In order to avoid it, we propose a mathematical trick. We propose to impose:

\begin{equation} \label{dumini}
\mbox{Domain}(\vec{q}_{i})\cap \mbox{Domain}(\vec{q}_{j})=\emptyset,
\end{equation}
and 

$$\mbox{Domain} (\vec{q}_{i})=V/N$$
for every $i$. That is: putting every particle in \textit{its own accessible volume}, which has the same magnitude for all of them. We do not imagine this solution as particles in pigeonholes, but only as a device for distinguishing them to count. In other words, we propose to substitute \eqref{plannn} by:
 
\begin{equation}  \label{pinxu} 
\varphi_{j} (\vec{p}_{j}, \vec{q}_{j})=\frac{1}{\left(\frac{V}{N} \right)^{1/2}} e^{i \frac{\vec{p}_{j} \vec{q}_{j}}{h}}.
\end{equation}
Note now, according to our procedure, we cannot write a global function analogue to \eqref{plannes}, because the domains of monoparticular functions are different (in fact, mutually exclusive). However, this trick allows us to write:

$$Z(T,V,N)=\left(\sum_{i} e^{-\beta \epsilon_{i}}  \right)^{N}=\left( \frac{V}{N \Lambda^{3}}\right)^{N}$$
($\Lambda$ is De Broglie thermal wavelength). Therefore, within the frame of \textit{Semicalssical Identity}, condition \eqref{helm} is trivially satisfied. In fact, this trick is mathematically equivalent to let the particles move along the whole volume (as it is usual) and then divide the resulting partition function by $N^N$. In sum, canonical partition function of a Maxwell-Boltzmann gas can be calculated because particles preserve their individuality, although they are not localized. Free energy is now certainly constituted of $N$ identical terms.

\section{Final Remarks}

We have tackled the question of identity with as much care as we have been able to do it. This old philosophical problem appears, in its physical mathematical form, as the problem of labelling identical particles. Starting from the Schr\"{o}dinger equation, we have analyzed two possible solutions for $N$-particle systems. 

 \textit{Quantum Identity}. The system is described through symmetrized or antisymmetrized eigenfunctions. According to this view, particles completely lose individuality. Put in a paradoxical way, \textit{they are so identical that they lose individuality}. This solution shows to what extent quantum mechanics casts a shadow on the concept of particle. However, the usual tendence of  keeping viewing $N$ as the number of particles evades the consequences of having symmetrized eigenfunctions: strictly speaking, we cannot individualize them when they are part of a gas, we cannot distinguish their particular momentum or energy.  This lack of localization has nothing to do with overlaping of individual eigenfunctions, a vestige of particle language. As we have seen, if we consider a gas in a volume $V$, we must symmetrize/antisymmetrize the eigenfunction to get rid of degeneration. But if we consider a container with a division in the middle ($V=V_{1}+V_{2}$), it is not necessary to symmetrize/antisymmetrize the eigenfunction of the whole volume, we must consider symmetrization only in every subvolume. As soon as we have constructed symmetrized/antisymmetrized eigenfunctions, we cannot keep on using particle images in the usual way: they have lost individuality, and there are not particular eigenfunctions which can overlap with the eigenfunction of another particle. Energy, momentum and position are the same for every particle. 
 
 Moreover, we have argued why we cannot even calculate Helmholtz free energy $F(T,V,N)$ of fermionic and bosonic ideal gases. It depends explicitly on the variable $N$ and it should be possible to decompose it in $N$, identical terms, but a fermionic or bosonic ideal gas cannot be conceived as a sum of $N$ identical parts.

In sum, we think it would be better not to keep talking about \textit{occupation numbers}, but about the \textit{number of monoparticular eigenfunctions which are included in the global eigenfunction} \eqref{psisa}. Another good example of how particle-picture remains is provided by the exclusion principle. It is said that \textit{two fermions cannot be in the same state}. On the contrary, we think that it should read: \textit{no monoparticular eigenfunction can appear twice in the global eigenfunction}. The usual formulation assumes the possibility of locating a fermion in a definite monoparticular state. But we have showed that this is not the case. And it is in view of that that we have avoided, as far as possible, using the expression \textit{indistinguishable particles}: it yields to imagining localized particles, when, as we have shown, that image is misleading. 

Within the \textit{Semiclassical Identity} approach we can keep talking of particles, but certainly they are \textit{indistinguishable} and non-localized. Their indistinguishability relies on the fact that different eigenfunctions \eqref{pitupitu} represent the same physical system, and therefore we cannot state which particle possess which momentum. This solution corresponds to Maxwell-Boltzmann statistics. Individuality is provided only by different momenta and requires complete absence of interaction. In this case, canonical partition function can be calculated through a mathematical trick consisting of assigning a subvolume $V/N$ to each particle. Only this way \textit{we can distinguish the indistinguishable} to perform the sum. Statistically, the system is completely equivalent to $N$ identical subsystems with the same (statistical) behaviour.

We opened our paper with an excerpt of Dirac. Let us close with another excerpt of a quantum physicist, Louis De Broglie. He noted many years ago the relation between interaction and lose of individuality \cite{brogl}:

\begin{quote}
C'est qu'en effet, une entit\'e physique \'el\'ementaire qui poss\'ederait l'autonomie individuelle dans toute sa pl\'enitude serait n\'ecessairement ind\'ependante de tout le reste de l'univers physique: petit monde ferm\'e, elle ne subirait aucune action et ne pourrait en exercer aucune. Pour pouvoir expliquer les ph\'enom\`enes à l'aide d'entit\'es \'el\'ementaires, il est doncs n\'ecessaire d'admettre qu'elles exercent entre elles des interactions: d\`es lors, ces entit\'es, \'etant en quelque mesure solidaires les unes des autres, ne seront plus aussi autonomes qu'on l'avait admis au d\'ebut et leur individualit\'e s'en trouvera quelque peu att\'enu\'ee. On con\c{c}oit alors combien int\'eressante du point de vue philosophique est la notion d'interaction parce qu'elle implique une certain limitation du concept d'individualit\'e physique.
\end{quote}

Complete loss of individuality or complete absence of interaction: these are the two alternative ways of dealing with identity we have found. The reflections by De Broglie about individuality in classical physics arose from the attempts of interpreting the new results coming from relativity and quantum mechanics. Those debates turned physicists back to the old mechanics to analyze concepts that were supposed to be well understood until then. But the new physics not always provides solid nor evident foundations for the old concepts. In fact, many times, it forces us to revisit them again and again.

\end{document}